\documentstyle[12pt, fleqn]{article}
\def\ref{\par\noindent\hangindent=6mm\hangafter=1}
\begin{document}
\baselineskip 8mm

\begin{center}

{\bf An Infrared Cutoff Revealed by the Two Years of {\it COBE}-DMR}\\

{\bf Observations of Cosmic Temperature Fluctuations}

\bigskip

Yi-Peng Jing$^1$ and Li-Zhi Fang$^{1,2}$\cite{byline}

$^1$ Department of Physics, University of Arizona, Tucson, 85721

$^2$ Steward Observatory, University of Arizona, Tucson, 85721


\end{center}

\bigskip
\bigskip

\begin{abstract}
We show that a good fitting to the first two years of
{\it COBE}-DMR observations of the two-point angular correlation
function of CBR temperature is given by models with a non-zero
infrared cutoff $k_{min}$ in the spectrum of the primordial density
perturbations. If this cutoff comes from the finiteness of the
universe, say, a topological T3 model, we find $k_{min} \sim
(0.3 - 1.1)\pi$H$_0/c$ with confidence level 95\%. Such a
non-zero $k_{min}$ universe would also give a better match
to the observations both of the RMS quadrupole anisotropy of CBR
and of galaxy clustering.
\end{abstract}



\newpage

Almost half a century ago, Infeld and Schild \cite{1} had pointed out that
if the size of the universe is finite, the infrared cutoff $k_{min}$
is then non-zero, and the infrared divergences in quantum field
theory are automatically excluded. However, it was shown later
that the problem of infrared divergence in QED can be solved even in an
infinite universe ($k_{min}=0$), because the infrared divergence in the
radiative correction can be precisely eliminated by adding the soft
photon bremsstrahlung to the vertex correction \cite{2}. In this
treatment we have, in fact, assumed $k_{min} < \Delta E/c\hbar$, where
$\Delta E$ is the energy resolution of the detection system.

For experiments done in physics laboratory the condition of
$k_{min} < \Delta E/c\hbar$ is always held. For instance, if the size
of the universe is of the order of today's horizon $c$H$^{-1}_{0}
\sim 3000 \ h^{-1}$ Mpc
(where $h$ is the Hubble constant in unit of 100 km s$^{-1}$Mpc$^{-1}$),
in order to detect soft photons with energy $\sim k_{min}$ the energy
resolution $\Delta E$ should be as small as $\hbar $H$_0$. The uncertainty
principle requires then that the time needed to measure the cosmic
infrared cutoff is $\sim$ H$_0^{-1}$. Therefore, it is impossible
to detect
the cosmic infrared cutoff $k_{min}$ by local experiments \cite{3}.

However, astrophysics do provide experiments, such as cosmic background
radiation (CBR), which have lasted as long as the age of the
universe, $\sim$ H$_0^{-1}$, and be able to measure $k_{min}$
comparable with H$_0/c$. Using this idea, we proposed that the
CBR anisotropy
is an effective tool to detect the size of a small universe. We
showed that the ratios of quadrupole moment of CBR anisotropy to higher
order multipoles sensitively depend on the ratio $L/(c$H$^{-1}_0)$,
where $L$ is the size of the universe \cite{4}. To compare this
$L/(c$H$^{-1}_0)$-dependence
with the first year data of the {\it COBE}-DMR \cite{5},
it has been found \cite{6} that the lower limit to the size of a cubic T3
universe
should be much larger than that given by the distributions of galaxies
and QSOs \cite{7}.

In this letter, we will show that a non-zero $k_{min}$ is possible. We are
motivated by the Two Years of {\it COBE}-DMR
observations of the CBR anisotropy (hereafter, the two-year data),
which has been available recently \cite{8}. Compared with the first year
data \cite{5}, the quality of the two-year data on many aspects
has been significantly improved. A new result is found to be that the RMS
quadrupole amplitude $Q_{rms}=6 \pm 3 \mu$K is significantly less than
the most likely quadrupole-normalized amplitude
$Q_{rms-PS} $, which is $12.4^{+5.2}_{-3.3} \mu$K for $n$=1.6, or
$17.4\pm 1.5 \mu$K for $n=1.0$, where $n$ is the index of the power-law
spectrum of density perturbation. This $Q_{rms}$-$Q_{rms-PS}$ difference
can not be totally explained by cosmic variance \cite{8}. On the other hand,
smaller $Q_{rms}$ may indicate the
lack of density fluctuations on the largest scales \cite{4}. Therefore,
it is valuable to study the possibility of explaining the
$Q_{rms}$-$Q_{rms-PS}$ difference by models with non-zero
infrared cutoff $k_{min}$.

Since the two-year data were
reduced in the scheme of standard inflationary universe ($\Omega =1$,
$\Lambda=0$, and simply connected spatial hypersurface), we
consider a Gaussian and adiabatic primordial perturbation with power-law
spectrum in a flat universe. The large-scale fluctuations in the CBR
temperature are given by \cite{9}:
\begin{equation}
{\Delta T\over T}({\bf \Omega})=-{H_0^2\over 2c^2}
\displaystyle\sum_{\bf k}{\delta ({\bf k})\over k^2}
e^{-i{\bf k}\cdot {\bf y}}\,,
\end{equation}
where ${\bf y}=(2cH^{-1}_0, {\bf \Omega})$ is a vector of length
$2cH^{-1}_0$ pointing to the direction ${\bf \Omega}$ on the
sky. $\delta ({\bf k})$ is the Fourier amplitude of the density
contrast $\delta({\bf r})$.
For a power-law perturbation, $\langle |\delta({\bf k})|^2
\rangle$ can be written as \cite{10}
\begin{equation}
\langle \mid \delta ({\bf k})\mid^2 \rangle
= {(2\pi)^3 \over V_\mu} \Delta^2_0{1 \over 4\pi} k^n \,,
\end{equation}
where $V_\mu$ is a large rectangular volume, $k=|{\bf k}|$
and $\Delta_0^2$ is a constant determined by the variance of
the perturbed potential.

The observed temperature fluctuations of CBR on the celestial
sphere are usually expressed by spherical harmonics
${\Delta T/T}({\bf \Omega})= \sum_{lm} a^m_l Y^m_l({\bf \Omega})$,
where $Y^{m}_{l} ({\bf \Omega})$ are the spherical harmonic functions.
Defining a rotationally invariant coefficient
$C_l\equiv \sum_m \langle |a^m_l|^2 \rangle$, one found from
eqs.(1) and (2) that
\begin{equation}
C_l={H_0^4(2l+1)\over 4 c^4}\Delta_0^2 {2\pi^2\over V_\mu}
\sum_{\bf k} k^{n-4} j_l^2(ky)\,,
\end{equation}
where $j_l(x)$ is the spherical Bessel function. Since our defined
$C_l$ is dimensionless, the {\it COBE} quadrupole amplitude $Q$ is
related to our quadrupole amplitude $C_2$ by $Q=C_2^{1/2}T$,
where $T$ is the mean temperature of CBR.

If the universe is of $k_{min}=0$, eq.(3) becomes \cite{11}
\begin{equation}
C^0_l={\frac {H^4_0(2l+1)}{4c^4}}\Delta^2_0
\int_0^{\infty} k^{n-2}j^2_l(ky)dk\,,
\end{equation}
where  the lower limit of the integration in eq.(4) is taken to
be zero, because $k_{min}=0$.
When $n<3$, eq.(4) gives
\begin{equation}
C^0_l={2l+1\over 5}C_2^0{\Gamma({9-n\over 2})\Gamma(l
+{n-1\over 2})\over \Gamma({n+3\over 2})\Gamma(l+{5-n\over 2})}\,,
\end{equation}
where $C^0_2$ is the quadrupole moment in a $k_{min}=0$ universe.

Let us consider a $k_{min} \neq 0$ universe, say, a cubic T3
universe \cite{4},  which is constructed from a
flat and infinite universe by the
following identification on the 3-dimensional flat hypersurface:
$(x_1+lL, x_2+mL, x_3+nL) =(x_1,x_2,x_3)$
for all integers $l,m,n$. $L$ is the size of the universe. In this case,
the coefficients $C_l$ should be directly calculated from eq.(3), i.e.
\begin{equation}
C_l={\frac {H_0^4(2l+1)}{16\pi c^4}} \Delta_0^2 y^{1-n}(yk_{min})^3
\sum_{\bf k} (ky)^{n-4} j_l^2(ky)\,,
\end{equation}
where $k_{nim}=2\pi/L$, and the summation covers all possible states
of the wave vector: ${\bf k}= k_{min}(l,m,n)$. Obviously, $C_l$ now
depends on three parameters: 1) the amplitude $\Delta_0$
of the perturbation (or the quadrupole moment $C_2^{1/2}$), 2) the infrared
cutoff $k_{min}$ (or the size of the universe $L$), and 3) the index $n$
of the perturbation spectrum. When $k_{min} \rightarrow 0$, or
$L \rightarrow \infty$, one has $C_l \rightarrow C^0_l$. Therefore,
this $k_{min}$-dependence of $C_l$ provides an effective method to detect
non-zero infrared cutoff $k_{min}$.

In the {\it COBE}-DMR observations, two measurements are independent:
a) the two-point angular correlation function $C(\theta)$ of CBR temperature,
and b) the RMS quadrupole amplitude $Q_{rms}$. The two-point angular
correlation function $C(\theta)$ is related to $C_l$ by
\begin{equation}
C(\theta)=\sum_l C_l W^2(l) P_l(\cos\theta)\,,
\end{equation}
where $P_l(x)$ is the Legendre function, and $W(l)$ is a window function,
which is  $W(l)=\exp\{ -1/2[l(l+l)/17.8^2] \}$ \cite{8}.
Comparing the measured correlation function to the model of eqs.(5)
and (7), one can find a most likely quadrupole-normalized amplitude
$Q_{rms-PS}$. The result based upon the first year data does show
$Q_{rms-PS} \sim Q_{rms}$ \cite{5}. Therefore, the model of $k_{min}=0$,
i.e. eqs.(4) and (5), is consistent with the first year data.
However, for the two year data, the difference between the RMS quadrupole
and the most likely normalized quadrupole is as large as $2 - 4 \sigma$
(Even taking the cosmic variance into account, the difference
is still significant at 90\% confidence level for $n=1$ \cite{8}).
Therefore, one would no longer be able to confidently say that the
standard model ($n=1$) is totally compatible with the current
{\it COBE}-DMR observations.

We use the standard $\chi^2$ technique to test the T3 models and to estimate
the most likely model parameters (quadrupole amplitude $C_2$ and
size $L$) by a $\chi^2$-minimization over
the two-point angular correlation function. For a given $n$ and
$y/L$, we estimate $C_{2\ rms-PS}$ by minimizing $\chi^2$ over the data:
\begin{equation}
\chi^2=\sum_i {[C_i-C(\theta_i)]^2\over
{\sigma_i^2+\sigma^2_{cv}(\theta_i)}},
\end{equation}
where $C_i$ and $\sigma_i$ are, respectively, the observed values and
errors of
the angular correlation at $\theta_i$, and
$\sigma_{cv}(\theta)$ is the $1\sigma$ cosmic variance of
the $C(\theta)$ \cite{12}.  In eq.(8) we assumed that the
variances in different bins are mutually independent. This approximation
is probably suitable for our purpose, because it has been known, at
least for the first year date, that the best fit amplitude of quadrupole
does not significantly depend on the nondiagonal part of the covariance
matrix of $C_i$ \cite{13}.
Since $\sigma_{cv}$ is also proportional to $C_{2\ rms-PS}$, we adopt
an iteration procedure to calculate $C_{2\ rms-PS}$. First we assume
a zero $\sigma_{cv}$ and find out
the best-fitting value of $C_{2\ rms-PS}$. Using this value we
calculate the $\sigma_{cv}$ based on 100 Monte Carlo realizations
of $C_l$, and do the minimization  again and find out a new fitting value
of $C_{2\ rms-PS}$.
Using this new $C_{2\ rms-PS}$ we repeat the minimization and find
another more accurate value of $C_{2\ rms-PS}$. The iteration procedure
is stopped until the differences of $\chi^2$ and
of $C_{2\ rms-PS}$ between the two consecutive minimizations
are less than 0.1\%. The final $C_{2\ rms-PS}$ and $\chi^2$ are
our desired values. In fact the calculation converges very fast,
and we need less than 5 iterations.
Fig.1 shows the goodness of the $\chi^2$-fit of each model, i.e.
the probability $P(> \chi^2_{min})$ that the experimental data are
drawn from a realization of the model.
It can be seen from Fig.1 that
for $k_{min} \sim 0$ models, i.e. $L \gg y=2cH^{-1}_0$,
the $P(>\chi^2_{min})$ of $n=1.6$ is much greater than that of
$n=0.6$ or 1.0.
This result is the same as that of likelihood analysis done by the
{\it COBE} group. Although the $n=1.6$ case can be comfortably
accommodated by the two-year
data [acceptance probability $P(>\chi^2_{min}) \sim 40\%$], the
smaller indices $n\le 1$, which are favored more by current
observational data of galaxy clustering \cite{14}, have
much lower acceptance probability.

Differing from the first year data, a remarkable feature shown
in Fig.1 is the high peak in
$P(>\chi^2_{min})$ space at $y/L \sim 0.8$.
This means that the best fit value of the infrared cutoff is
$k_{min} \sim 0.8\pi H_0/c$. In Fig.2 we plot such a  best-fitting
curve as well as
the two-year data of $C(\theta)$. We also plot
the best-fitting curve of an $k_{min}=0$ universe ($n=1.6$)
in Fig.2. It can be clearly seen from both Figs.1 and 2 that
considering the possible existence of a non-zero cosmic
infrared-cutoff substantially improves the model fit to the
experimental data (even for $n=1.6$), and the most probable
value of $y/L$ is $\sim$0.8, which is almost independent of
$n$. At the 95\% confidence
level, we have $0.3<y/L<1.1$ for $n=1$, $0.5<y/L<1.1$ for
$n=0.6$ and $y/L<1$ for $n=1.6$.

Fig.3 plots our best-fitting $C_{2\ rms-PS}^{1/2}$ as a function
of $y/L$, where the index $n$ is taken to be 0.6, 1.0
and 1.6 as well. We plot the measured value of $C_{2\ rms}^{1/2}$
by the bold line in Fig.3, and its $1\sigma$ range by the dotted
area. One can find that the models with
$L < 1.2y=2.4 cH^{-1}_0$ give good agreement between
$Q_{rms-PS}$ and $Q_{rms}$, i.e. their difference is no longer
larger than 1 $\sigma$. This result is also almost independent of
the power law index $n$.

Because a finite universe lacks the fluctuation power on scales
larger than its size, we require a larger amplitude $\Delta_0$
of the density perturbation to fit with the observational
$C(\theta)$ \cite{15}. However, larger $\Delta_0$ will lead to a
stronger clustering on smaller scales. Therefore we should study
whether the amplitude
$\Delta^2_0$ in a $k_{min} \neq 0$ universe is compatible with
observed clustering
of galaxies on scales, say, $8 \,h^{-1}$Mpc. In order to calculate
the density
fluctuations on smaller scales from a given $\Delta_0$ perturbation,
we need to choose the transfer function $T(k)$ of linear growth \cite{9}.
The $T(k)$ in turn is mainly determined by the matter composition of the
universe. Here we use the $T(k)$ of the standard cold dark matter (CDM)
model and of the cold plus hot dark matter (CHDM) model
($\Omega_{CDM}=0.7$, $\Omega_{\nu}=0.3$, and
$h=0.5$ \cite{16}). Fig.4 presents the predicted values of $\sigma_8$,
the $rms$ density fluctuation of a sphere $r=8 \,h^{-1}$Mpc,
 as a function of $y/L$ for $n=1$. In Fig.4 we also plot
a result of $\sigma_8$ given by statistics of galaxy distribution.
By examining clusters of galaxies, White et al. \cite{17} suggest that
in an Einstein-de Sitter universe
$\sigma_8$ is between 0.52 and 0.62. These limits
are shown as the solid lines in Fig.4. Obviously, small universes
of $y/L\gg 1$ are not acceptable, because they produce excessive clustering
on $8 \,h^{-1}$Mpc scale for both $T(k)$.
The models of $y/L \leq 1$ are
acceptable when the CHDM $T(k)$ is used.
[Considering uncertainties in the
estimated limits of $\sigma_8$ and in the choice of $T(k)$, we caution
readers against placing overaccurate constraints of the $\sigma_8$-test
on the universe sizes].

With reasonable transfer functions, the best-fitting value of
power index $n=1.6$ for large
universe ($L \gg 2cH^{-1}_0$) is hard to reconcile with the observational
data of galaxy clustering \cite{14}.
While the $COBE$-DMR data allows the finite universe of
$L \sim 1.2 cH^{-1}_0$ to have $n=1$, which can be brought into agreement
with galaxy clustering for some reasonable transfer functions.
Therefore, compared with models of $k_{min}=0$ (infinite) and
large $k_{min}$ (small)
universes, the moderate-size (or $k_{min} \sim \pi$H$_0/c$) universe is
in better agreement with the observations
when both the $COBE$-DMR result and the clustering of
galaxies are considered.

In summary, in terms of the three independent and basic tests, i.e.
$C(\theta)$, $Q_{rms}$  and the galaxy clustering, a good survived
model among those discussed in the paper is the one of a non-zero
$k_{min}$ universe. If this cutoff comes from the
finiteness of the universe, such as a topological T3 model, the most
probable value of $k_{min}$ is $0.8\pi H_0/c$. However, it does not mean
that the $k_{min} \neq 0$ can only be given by a
multiply connected topology like T3. There are other mechanism for
$k_{min} \neq 0$.  For instance, in the standard
inflationary universe, if the parameter $N_{tot}$ in the inflation
factor $\exp(N_{tot})$ is $\sim 54$ (somewhat of fine-tuning parameter
may be needed), one would also have an initial density perturbation
with spectrum cutted off at $k_{min} \sim \pi H_0/c$.
Regardless of these variousness of
possible explanations, one can conclude, at least, that the {\it COBE}-DMR
observation of $\Delta T/T$ is a powerful tool to probe the cosmic
infrared cutoff. The first two years of the {\it COBE}-DMR observations
reveal the possible
existence of non-zero infrared cutoff in the spectrum of the primordial
density fluctuations. It seems to be the first time to derive an
"observed" values of the cosmic $k_{min}$ with confidence higher than
$95\%$. However,
we should remember the difficulty in the measurement of $Q_{rms}$
due to foreground contaminations. Therefore, in order to discriminate
various explanations, it is necessary to have other tests.
The amplitude of octavopole $C_3$ of $\Delta T/T$ would be one of such tests
 \cite{18}. We believe, more convinced results
related to the non-zero cosmic infrared cutoff will be obtained as more
observations of CBR $\Delta T/T$ become available.

\bigskip

YPJ thanks the World Laboratory for a Scholarship. The two-year data was
kindly provided by C.L. Bennett. The work is
partially supported by NSF contract INT 9301805.


\newpage



\newpage

{\bf Figure captions}

\bigskip


Fig.1 $P(>\chi^2_{min})$ as a function of $y/L$. Here $\chi^2_{min}$
is calculated by fitting the finite $L$ universe model to the
{\it COBE}-DMR two-point angular correlation function
of the cosmic temperature fluctuations. The index $n$ of power law
perturbations is taken to be 0.6, 1.0 and 1.6. The dot-dash line
denotes $P(>\chi^2_{min})=5\%$.



Fig.2 The observed angular correlation $C_l(\theta)T^2$ and the
best-fit curves by a) a T3 model with $n=1$ and $y/L=0.80$ (solid line),
and b) an infinite and flat universe with $n=1.6$ (dashed line).



Fig.3 The best fitted quadrupole, i.e. $C_{2\ rms-PS}^{1/2}$, as a
function of the size of the universe. The thick line denotes the RMS
quadrupole measurement $C_{2\ rms}^{1/2}$, and the dotted area is its
$1\sigma$ region.



Fig.4 The predicted $\sigma_8$ in a cubic T3 models. Two types of
transfer functions are assumed: the standard CDM (dotted line) and
the CHDM (dashed line). The solid lines are the upper and lower limits
to $\sigma_8$, given by statistics of galaxy distribution [17].



\begin{thebibliography}{99}
\bibitem{15} It is easy to see from eq.(7) that the low-order $C_l$
contributes significantly to the angular correlation function
$C(\theta)$ at small $\theta$. Since in T3 model low-order $C_l$ is
smaller, it requires a higher amplitude $\Delta_0$ in comparing
with a $k_{min}=0$ universe.
\end{thebibliography}
\end{document}